\documentclass{article}
\usepackage[utf8]{inputenc}
\usepackage{url}
\usepackage{geometry}\providecommand{\keywords}[1]
{
  \small	
  \textbf{\textit{Keywords---}} #1
}

\usepackage[justification=centering]{caption}
\usepackage{amsmath}
\usepackage[table, dvipsnames]{xcolor}
\usepackage{amssymb}

\usepackage{mathptmx}       
\usepackage{helvet}         
\usepackage{courier}        
\usepackage{type1cm}        
\usepackage{makeidx}         
\usepackage{graphicx}        
\usepackage{tabularx}
\usepackage{float}
\usepackage{subcaption}
\usepackage{multirow}
\usepackage{textcomp}
\usepackage[linesnumbered,ruled]{algorithm2e}
\usepackage{multicol}        
\usepackage[bottom]{footmisc}

\title{HBFL: A Hierarchical Blockchain-based Federated Learning Framework for a Collaborative IoT Intrusion Detection}
\author{Mohanad Sarhan*$^{1}$, Wai Weng Lo$^{1}$, Siamak Layeghy$^{1}$,  Marius Portmann$^{1}$ \\
        \small $^{1}$University of Queensland, Brisbane, Australia \\
        \small *Corresponding Author: m.sarhan@uq.net.au\\
}

\date{}

\begin{document}

\maketitle

\begin{abstract}
The continuous strengthening of the security posture of IoT ecosystems is vital due to the increasing number of interconnected devices and the volume of sensitive data shared. The utilisation of Machine Learning (ML) capabilities in the defence against IoT cyber attacks has many potential benefits. However, the currently proposed frameworks do not consider data privacy, secure architectures, and/or scalable deployments of IoT ecosystems. In this paper, we propose a hierarchical blockchain-based federated learning framework to enable secure and privacy-preserved collaborative IoT intrusion detection. We highlight and demonstrate the importance of sharing cyber threat intelligence among inter-organisational IoT networks to improve the model's detection capabilities. The proposed ML-based intrusion detection framework follows a hierarchical federated learning architecture to ensure the privacy of the learning process and organisational data. The transactions (model updates) and processes will run on a secure immutable ledger, and the conformance of executed tasks will be verified by the smart contract. We have tested our solution and demonstrated its feasibility by implementing it and evaluating the intrusion detection performance using a key IoT data set. The outcome is a securely designed ML-based intrusion detection system capable of detecting a wide range of malicious activities while preserving data privacy.
\end{abstract} \hspace{10pt}


\section{Introduction}

IoT (Internet of Things) is an ecosystem of interconnected devices, each embedded with computational tools, such as sensors or processing units, to collect, store, and exchange data over the internet \cite{lee2015internet}. Each device performs a task to provide the overall function or purpose of the IoT ecosystem, such as smart homes, healthcare, or industrial operations. IoT enables remotely controlled devices to collaborate and achieve common objectives. According to research by the International Data Corporation (IDC), due to the demand for IoT networks, it is estimated that the number of IoT devices would reach up to 41.6 billion in 2025 \cite{wu2020convergence}. IoT ecosystem can offer great potential to businesses and consumers due to its digital and automated intelligence capabilities, increasing operational agility and efficiency. However, the security of IoT networks has become one of the main challenges in the implementation of such networks \cite{zhang2014iot}. 

IoT devices collect and store highly sensitive data, such as personal, financial, and medical information, which are often the target of hackers \cite{manworren2016you}. The current state of IoT networks' security posture makes it harder for businesses and consumers to preserve and trust the security of their digital assets. This is often due to the large, complex, and widely distributed attack surface as IoT devices are often located in the edge zone of the perimeter \cite{ghirardello2018cyber} and provide an entry point into an organisational core network. These devices are generally interconnected and in constant communication, forming an IoT ecosystem; therefore, infiltration of any of them poses serious concerns for the privacy and security of the entire network. Furthermore, since the IoT connects the digital and physical worlds, hacking into IoT devices can cause critical concerns \cite{alaba2017internet}, such as the hijacking of the Ukrainian power grid in 2015 \cite{case2016analysis}, causing power loss in a city with a population of around 1.4 million. 

Therefore, Machine Learning (ML) capabilities have been endorsed in the development of IoT Intrusion Detection Systems (IDSs) to enhance the validity of transmitted data against cyber attacks. ML is widely and successfully integrated into many domains to empower decision-making systems. The development and usage of ML-based IDS have attracted great attention in the cyber security field, to preserve the three principles of information systems; confidentiality, integrity, and availability \cite{stair2020principles}. Current traditional detection methods, such as signature-based IDS, are highly dependent on IOC (Indicators of Compromise) to discover malicious activities. However, ML-based IDSs have been designed to distinguish between benign and malicious computer activities using behavioural patterns. Therefore, the intelligent detection method is highly effective in the discovery of zero-day and Advanced Persistent Threats (APTs) compared to traditional detection methods \cite{ahmad2021network}.

The collaboration between multiple organisations/entities to share Cyber Threat Intelligence (CTI) to improve the performance of IDSs is not uncommon in the cyber security field \cite{barnum2012standardizing}. Several sources publicly share a large number of IOCs to be used in the monitoring and detection of emerging threats by signature-based IDS. The sharing of such IOCs does not represent any security or privacy risks to any organisation. However, as ML-based IDSs do not require the registration of IOCs, but rather patterns extracted directly from benign and malicious data, CTI has been difficult to apply in such systems \cite{conti2018cyber}, mostly because of security and privacy concerns associated with data sharing. Although CTI can accelerate the learning process and improve the attack detection capabilities of the learning model, there is a limited amount of research work providing a solution enabling CTI between organisations adopting ML-based IDS. 

In this paper, we propose a Hierarchical Blockchain-based Federated Learning (HBFL) framework to enable CTI between organisations adopting ML-based IDSs in IoT ecosystems. The outcome of HBFL is a group of organisations benefiting from each other's threat intelligence while maintaining the privacy of sensitive training log sets internally. HBFL will promote the practical deployment of ML-based IDSs and improve their attack detection accuracy. HBFL follows a federated learning scenario, which has been developed to retain the same benefits of centralised learning methods without the need for local data exchange \cite{li2020federated}. Data used in ML training often contain highly sensitive information about the users and activities of the IoT network. Therefore, the security preservation of the IoT endpoints and their data privacy is critical. Federated learning allows training a common ML model across multiple endpoints locally, thus preserving data privacy, which is critical for IoT ecosystems and the key motivation for the development and usage of federated learning in general.

The HBFL model is exposed to a variety of heterogeneous data sources and types available at a wide range of IoT endpoints \cite{bouacida2021vulnerabilities}. Therefore, it increases the generalisability and detection performance of ML-based IDS across several attack types and benign usage scenarios. Moreover, a vertically integrated variant of federated learning called Hierarchical Federated Learning (HFL) \cite{liu2020client} has been used to enforce additional privacy-preserving mechanisms and accommodate large-scale deployments. The key difference is the addition of a middle layer of servers (combiners) performing model weight aggregation \cite{abad2020hierarchical} between the group of participating organisational IoT endpoints and the global server (reducer). These additional control point mechanisms attract many benefits in the management and governance of the HBFL framework and provide defence layers against adversarial attacks \cite{9127823}. 

HBFL removes the trust requirement between organisations or a centralised third party to orchestrate and maintain the framework functionality. Due to embedded smart contract technologies performing the reducer roles in the federated learning operation. Smart contract overcomes the lack of trust behaviour which is often a roadblock in any collaboration process and will motivate and assist the participation of organizations. HBFL orchestration is performed independently, securely, and transparently on a permissioned blockchain system that is monitored and validated by all participants. The smart contract will contain the planned HBFL tasks and will monitor the execution results to ensure that only conforming tasks advance the workflow. The permissioned blockchain network provides a trusted hosting platform for transactional records and values, enabling auditing and accounting for any anomalies or fraud.

In this paper, we propose a novel HBFL framework for collaborative detection of IoT intrusions. The framework adopts a cloud-fog-edge topology in which IoT endpoints and combiners are hosted on edge and fog perimeters, respectively. The reducer runs on permissioned blockchain-based cloud computing. Furthermore, the overall HFL process will be orchestrated by a smart contract running on the blockchain. The benefits of the proposed HBFL framework are;
\begin{itemize}
    \item providing a secure solution to enable CTI between organisations, 
    \item maintaining the privacy of local IoT data logs, 
    \item scalable framework to allow real-world design and deployment, 
    \item omitting the requirement of trusting a single entity to orchestrate the federated learning process, and
    \item logging the transactional details in a private immutable ledger.
\end{itemize}

The outcome of HBFL is a securely designed collaborative IoT-IDS model reflecting the global intelligence of several endpoints while ensuring the privacy and security of sensitive training log sets. Most importantly, the framework solves real-world challenges, such as sharing data and managing processes workflow, increasing the chances of potential deployment in the commercial space. The framework has been evaluated using a key IDS data set, and the results illustrate the feasibility and efficiency of HBFL. The following section provides background and motivation for the technologies adopted in this paper, followed by a brief review of the related work in Section \ref{rw}. The design, components and functionality of HBFL are explained in Section \ref{fw}. Furthermore, Section \ref{em} provides an evaluation use case study in which a prototype is developed, evaluated, and compared with non-collaborative scenarios. The results are provided and discussed using various evaluation metrics before concluding.

\section{Background}
\label{bg}
In this section, the motivation of HBFL i.e. CTI and its primary components, i.e. federated learning and blockchains, are introduced and discussed.

\subsection{Cyber Threat Intelligence}
CTI is one of the widely adopted techniques in the defence community against cyber attacks. The ultimate goal of CTI is to share the experiences faced by an organisation, this includes the attack types encountered, as well as the techniques and tactics used in their execution \cite{mavroeidis2017cyber}.   Many organisations using signature-based IDS heavily rely on shared intel, such as Malware Information Sharing Platform (MISP) \cite{wagner2016misp}, an open-source CTI platform. Shared intelligence is found in different forms such as documentation, IP addresses, hash values, and domain names associated with external threat actors. This intelligence is directly fed to signature-based IDS for analysis and detection. 

The collaborative CTI sharing enables each participant to learn from experiences encountered by other organisations \cite{zhao2018federated}. In ML-based IDS, CTI is challenging to achieve, as benign and malicious training data sets from several organisations are required for model training, which often poses privacy and security concerns. More importantly, recently introduced laws such as the General Data Protection Regulation (GDPR) \cite{truong2021privacy} in Europe aim to protect data privacy and address unauthorised sharing of sensitive user information. Infraction of such laws often attracts serious legal issues and hefty fines, in the case of GDPR, the fine can be up to \$20 million \cite{8190804}. 

However, without CTI, the organisational defence model is limited to local experience and might not generalise across non-Independent and Identically Distributed (IID) benign and attack sources. Therefore, current methods for developing IoT IDSs do not scale and are vulnerable to the rapid growth of IoT ecosystems due to the introduction of new devices, services, and emerging zero-day attacks. Moreover, the requirement of collecting a large amount of corresponding training data samples for each data class is often a major barrier for small organisations aiming to effectively design an ML-based IDS, which is addressed by HBFL. Therefore, our proposed HBFL framework aims to enable CTI between organisations adopting an ML-based IDS. HBFL develops a more comprehensive IDS model by gathering data from several organisations to be used in the training process.

\subsection{Federated Learning}

ML models are statistical algorithms capable of learning complex patterns from historical data to provide intelligent insights into future data \cite{surden2014machine}. The design of such ML models requires a training process in which the learning models extract useful semantics from the available data logs.  In an IoT ecosystem, these training log sets are generated locally by many different digital endpoints such as mobile phones, smart appliances, sensors, etc. The quality and quantity of data used in the training and evaluation of learning models have a direct effect on the performance of ML-based IDSs \cite{abdelmoumin2021performance}. There are three main types of ML; a) localised learning - the model is trained locally using data sets generated by a single endpoint \cite{taherkhani2016centralized}, b) centralised learning - the model is trained centrally using data sets collected from multiple endpoints \cite{taherkhani2016centralized}, and c) federated learning - the model is trained across multiple endpoints using local data sets without exchanging them \cite{yang2019federated}. 

\begin{figure*}[h!]
\centering
\begin{subfigure}{.2\textwidth}
  \centering
  \includegraphics[width=3cm, height=2cm]{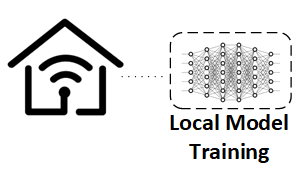} 
  \caption{Localised}
\end{subfigure}
\hfill
\begin{subfigure}{.3\textwidth}
  \centering
  \includegraphics[width=5cm, height=4.5cm]{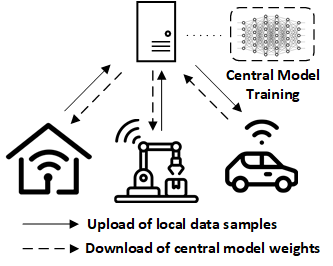}
  \caption{Centralised}
\end{subfigure}
\hfill
\begin{subfigure}{.4\textwidth}
  \centering
  \includegraphics[width=7cm, height=4.5cm]{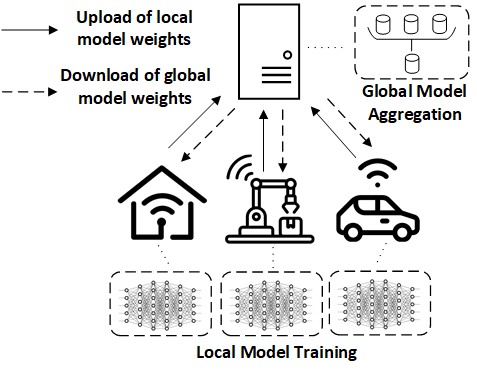} 
  \caption{Federated}
\end{subfigure}
\caption{Machine learning scenarios}
\label{mm}
\end{figure*}

There are benefits and limitations to each of the different ML scenarios. Figure \ref{mm} represents the architectural difference between each ML scenario. Localised ML models often achieve great performance over local IID data samples \cite{zhang2021survey}. However, this performance does not generalise across other non-IID and heterogeneous data samples. Given that the network and host data are unique to each endpoint's activity, localised learning is unfeasible in the design of ML-based IDSs \cite{sarhan2021cyber}. Therefore, centralised ML models were designed to overcome such limitations by training over multiple sourced non-IID data samples. This exposes the learning model to a large number of heterogeneous data samples, each representing a unique behaviour of benign and malicious logs. However, this method does not consider security aspects or respect users' privacy, due to the requirement of collecting and sharing logs centrally \cite{tedeschi2021iotrace}. This often exposes sensitive information and behaviour in the case of IoT data. 

Federated learning techniques have been developed to ensure the privacy of data used in the design of ML by training a model locally at each participating endpoint \cite{yang2019federated}. Only the updated model parameters are shared with a trusted entity where an aggregation process is performed to combine local updates received from several endpoints into a single global model \cite{mcmahan2017communication}. The key benefit of adopting a federated learning technique in the design of ML-based IDS is the exposure of the model to a variety of heterogeneous data sources. This method improves the generalisability and detection performance of IoT attacks while protecting the privacy of local endpoint data samples \cite{li2020federated}. Moreover, the resource requirements such as storage, computational power, and network bandwidth are reduced as there is no transfer or collection of training data sets centrally. Therefore, the application of federated learning in IoT networks is highly motivated, and many use cases have been proposed in the literature that adopt this architecture, including the design of IoT IDSs \cite{popoola2021federated}.

The most common federated learning aggregation technique is proposed by Google, known as Federated Averaging (FedAvg) \cite{mcmahan2017communication}. The weight averaging algorithm attracts many benefits such as being robust to non-IID and unbalanced data distributions. It is also cost-efficient and can run on endpoints with limited communication capabilities by reducing the number of federated learning rounds required to reach maximum performance. In a standard federated learning scenario using FedAvg, the global server broadcasts an initialised model weight $w$ to each client $k$. Each client computes the gradient $g_{k}$ on its local training data samples. The updated parameters are calculated using $w_{t+1}^{k} \leftarrow w_{t}-\eta g_{k}$ for $E$ epochs, where $t$ represents the federated learning round and $\eta$ is the local learning rate. Once the required number of updated weights is received during the round $w_t$, the global server aggregates them using Equation \ref{a}, where $n$ is the number of participants.

\begin{equation}
w_{t+1} \leftarrow  \sum_{k=1}^{K} \frac{n_{k}}{n} w_{t+1}^{k}
\label{a}
\end{equation}

However, traditional federated learning has raised various challenges and limitations in real-world deployments \cite{abad2020hierarchical}. This includes the extreme dependence on a single global server to orchestrate the federated learning process, including the aggregation of model updates. Therefore, the scalability of federated learning is often questionable, as a large number of different clients with unique model updates are connected to and aggregated by one global server \cite{abad2020hierarchical}. This method might be unfeasible in wide-scale deployments such as industrial IoT ecosystems due to communication links and scarcity of computational resources. Moreover, centralisation of the global server introduces a single-point-of-failure concern in the federated learning operation. Although this limitation has been partially addressed by the hosting of the global server on distributed de-centralised networks such as blockchains \cite{9127823}, other key challenges are still faced. This includes the recipient of model updates directly from user endpoints, which can be re-engineered to raw data using advanced adversarial attack techniques \cite{bhagoji2019analyzing}. Additionally, any poisonous updates received from an adversary will have a direct impact on the entire federated learning process. In this paper, we explore a hierarchical federated learning architecture to address these drawbacks.

\subsection{Blockchain}

Blockchains are fundamentally an electronic ledger of digital transactions that are duplicated and hosted by a distributed and de-centralised network of computer nodes \cite{christidis2016blockchains}. Blockchain provides a trusted storage platform for transactional records and values. Participants share transactional data in a large network of untrusted nodes that are resistant to modification or deletion using cryptographic proofs \cite{christidis2016blockchains}. Thus, making it a reliable source for future accounting and auditing purposes. A blockchain is characteristically made up of two parts; transactions and blocks \cite{tasca2017taxonomy}. Transactions are defined as the content created by a user who wants to store information on the blockchain. Blocks are groups of transactional records that confirm the sequence and integrity of transactions that have occurred. The block is then hashed and kept as a record that is available to all entities, explaining the terms of a transparent ledger.

A type of blockchain with managed Role-Based Access Control (RBAC) is known as a permissioned blockchain \cite{helliar2020permissionless}. Unlike private blockchains, permissioned blockchains are decentralised across multiple nodes. They are used mainly by organisations to manage supply chains and other collaborative processes where security and identity management are required. The network is owned by a single or a group of predefined nodes known as administrators who manage the RBAC. The permissioned ledger is not publicly accessible and can be accessed by users who can only perform specific actions granted to them by an administrator \cite{helliar2020permissionless}. The users identify themselves via digital certificates, signatures, or credentials. New users must be invited to access the network. In addition, they are generally more efficient compared to public blockchains due to the limited number of nodes on the network, hence increasing the transactional rate and overall performance.

Blockchains provide a computational infrastructure capable of running programmes known as smart contracts. Smart contracts are arbitrary user-defined contract programs that can execute the terms and conditions of a contract \cite{zheng2020overview}. The general objectives are to satisfy contractual conditions, minimise errors, and eliminate the need for trusted intermediaries. They are programmed to monitor, log, and execute shared tasks. These smart contracts act as a trusted shared agreement that contains all planned interactions between the parties involved \cite{zheng2020overview}. Smart contracts aim to help organisations collaborate without the need to trust the parties involved or a third party mediator. Lack of trust is often a roadblock leading to conflicts that stall the overall collaboration process, such as which party will handle the orchestration process in federated learning. Moreover, the dependency on one another to perform within the expected quality guidelines or timeline usually leads to inflictive arguments once unexpected kinds of issues arise.

\section{Related Work}
\label{rw}
In this section, we discuss and analyse some of the key related works found in the literature. In Cui et al. \cite{cui2021security}, the authors proposed an anomaly detection framework for IoT infrastructures. The framework is empowered by a public blockchain and adopts a federated learning architecture for the ML operation. The blockchain network ensures data integrity and improves efficiency while being robust to server staleness or downtime. The federated learning technique uses a Generative Adversarial Network (GAN) neural model to further enhance the privacy of local model parameters by injecting controlled noise into the updates received, also known as differential privacy. The framework was evaluated on a single benchmark data set, and the results demonstrate high accuracy of anomaly detection and improved privacy protection. The data set used is the well-known KDD-CUP dataset, which was released in 1999; therefore, it does not represent the modern attack behaviour of IoT \cite{8672520}.

Lu et al. \cite{8843900} integrated a federated learning operation with a blockchain network to address the current limitations of data sharing mechanisms in IoT ecosystems. The proposed framework preserves the privacy and security of information by adopting a permissioned blockchain where data owners can further control access to the shared updates of the ML model. When a new endpoint participates in the ML operation, its unique identity is recorded on the blockchain as a transaction together with the profile of its data, including data categories, type, and size, which will be verified by each node. The participants are divided into segments according to their data types and categories. The framework is evaluated using two benchmark data sets for data categorisation. The results demonstrate the efficiency of data classification while improving privacy. However, the increase in data providers brings more local models to be updated and computed, increasing the aggregation time.

In \cite{demertzis2021blockchained}, an intelligent threat defence system for smart cities has been developed. The framework employs blockchain and federated learning technologies to ensure the privacy and anonymity of participants. The learning process is achieved through encrypted smart contracts within the blockchain network for unambiguous validation and control of the process. The proposed system aims to detect intrusions in IoT traffic using deep content inspection. The system has been designed using five main components; 1) blockchain server - the consensus mechanism used for validating the transactions, 2) public key server - to create public and private keys for user identity validation, 3) federated learning server - controls the execution of the distributed learning, 4) miner - hosted on the endpoints to send updates of the local model and detect updates of the global model, and 5) machine learning server - contains the learning model to be trained using local data. The framework does not appear to be evaluated or validated in the paper.

Liu et al. \cite{liu2021blockchain} aimed to protect vehicular networks and modern transportation infrastructure. Due to the wide autonomous vehicle attack surface and the expanded use of software and wireless interfaces, current IDSs can be customised efficiently. The article integrates efficient model training and secure model sharing using federated learning and blockchain technologies, respectively, to ensure driving safety. The public Ethereum blockchain network is used for the storage and sharing of training models. The design of the proposed framework involves a two-stage IDS that leverages federated learning through several edge vehicles and roadside facilities. The aggregation layer is replaced by distributed roadside units. The uploaded model is combined with mask noise based on secret sharing to ensure the privacy of user data. The work also discusses a trust-based incentive mechanism to achieve a high detection rate. The framework is evaluated using the outdated KDDCup99 data set, which does not represent modern IoT attack behaviour \cite{8672520}. A feed-forward artificial neural network is used to perform the classification task. The proposed scheme achieved a reliable accuracy score through aggregation training.

Zhang et al. \cite{9233457} have designed a failure detection system for industrial IoT networks. The architecture involves blockchain-based federated learning, where each client periodically creates a hashed data record on a blockchain. The paper also developed a novel aggregation technique known as Centroid Distance Weighted Federated Averaging (CDW\_FedAvg), which takes into account the distance between positive classes and negative classes of each client data set. In addition, a smart contract is developed to reward each client according to the size and centroid distance of the data used in model training. The framework is evaluated using a real-world industry partner data set, one of the largest air conditioner manufacturers in the world. In the experiment, there is a central organisation which is the platform owner (manufacturer) that maintains the industrial services and there is a client organisation which is in this case hotels that contribute with local data and computation resources. The performance has been compared to the traditional FedAvg technique \cite{mcmahan2017communication} and the results demonstrate the superiority of the CDW\_FedAvg technique in this use case.

Chai et al. \cite{9127823} proposed a blockchain-based hierarchical federated learning algorithm for knowledge sharing among intelligent vehicles. The authors highlight the importance of knowledge sharing across vehicles and how it improves the decision capabilities of the model. Privacy, security, and reliability issues in knowledge sharing can disturb the sharing system.  Due to the hierarchical architecture, the framework is feasible for large-scale vehicular networks and can effectively reduce computational consumption. The process follows a game theory in which the multi-leader and multi-player trade parameters are in the trading market. The framework is evaluated using the MNIST and CIFAR10 data sets. The simulation results show that it can improve sharing efficiency and learning quality by about 10\% compared to traditional federated learning methods.

Saadat et al. \cite{9460304} evaluated and compared the performance benefits of HFL with standard federated learning techniques in IoT intrusion detection applications. In particular, the authors measured the detection accuracy and speed of convergence of each method using a neural network model evaluated on the outdated NSL-KDD data set samples. The paper explored three study cases with different client-edge assignments to observe the effect of non-IID data distribution over the clients. The data set was divided according to attack classes; the first use case provided IID data assignments for federated learning and HFL techniques. The second use case provided IID data assignments with an equal number of attack samples. The third use case provided non-IID data assignments. The results proved HFL's superiority over federated learning in the three study cases, where HFL achieved a faster convergence rate, lower loss, and higher testing accuracy.

In Preuveneers et al. \cite{app8122663}, the authors have evaluated the feasibility of federated learning based on permissioned blockchain for intrusion detection. The transactional records in this case are weight updates to the ML model, which are monitored for tamper attempts or fraudulent transactions of adversaries. The framework used a permissioned blockchain network known as MultiChain and is evaluated using a key IDS data set known as CICIDS2017. In the training process, a one-class classification technique was followed using an auto-encoder model. The goal of this work is to measure the computational impact of blockchains in a federated learning scenario. The experiments illustrate that the increased complexity caused by blockchain technology has a limited performance impact (between 5\% and 15 \%) on federated learning. The development of smart contracts is highlighted as future work.

Sarhan et al. \cite{sarhan2021cyber} aimed to provide a CTI enabling solution for organisations adopting ML-based IDSs. The paper argues the benefits of such collaboration and the value added to the detection capabilities of the model. The proposed framework follows a federated learning architecture where each organisation centrally collects the training log sets and trains an ML model locally. The updated parameters are received by the global server, which is aggregated with other organisations' locally updated weights. The framework is evaluated using two key NetFlow IDS data sets; NF-UNSW-NB15-v2 and NF-BoT-IoT-v2 used together in parallel during training. The benefits and efficiency of collaboration are demonstrated by comparing the detection results with localised and centralised ML scenarios. Using two widely used ML neural network models, the results of the federated learning scenario are superior to the localised learning scenario and slightly inferior to the centralised learning scenario. However, data privacy is only preserved using federated learning.

\begin{table}[ht]\footnotesize
\centering
\caption{Related works}
\begin{tabular}{|l|l|l|l|l|l|l|}
\hline
\textbf{Paper}              & \textbf{Fed. Learning}          & \textbf{Blockchain}   & \textbf{Smart Contract} & \textbf{IoT} & \textbf{IDS} & \textbf{CTI} \\ \hline
Cui et al. \cite{cui2021security}        & Standard             & Public                & No                     & Yes          & Yes  & No        \\ \hline
Lu et al.  \cite{8843900}        & Standard             & Permissioned          & No                     & Yes          & No   & N/A        \\ \hline
Demertzis \cite{demertzis2021blockchained}        & Standard             & Public                & Yes                    & Yes          & Yes  & No      \\ \hline
Liu et al. \cite{liu2021blockchain}       & Standard             & Public                & No                     & Yes          & Yes  & No       \\ \hline
Zhang et al. \cite{9233457}     & Standard             & Public                & Yes                    & Yes          & No   & N/A       \\ \hline

Chai et al. \cite{9127823}      & Hierarchical          & Public                & No                     & Yes          & No & N/A           \\ \hline
Saadat et al. \cite{9460304}       & Hierarchical          & No                & N/A                     & Yes          & Yes & No           \\ \hline
Preuveneers et al. \cite{app8122663}& Standard             & Permissioned          & No                     & No           & Yes  & No        \\ \hline
Sarhan et al.  \cite{sarhan2021cyber}    & Standard             & No                    & N/A                    & No          & Yes  & Yes      \\ \hline 
\textbf{This Paper}         & \textbf{Hierarchical} & \textbf{Permissioned} & \textbf{Yes}            & \textbf{Yes} & \textbf{Yes} & \textbf{Yes} \\ \hline
\end{tabular}%

\label{t-rw}
\end{table}

Significant research has been conducted on the integration of blockchain and intrusion detection to improve data privacy and attack detection, respectively. Table \ref{t-rw} provides a snapshot of the articles mentioned above, listing the types of federated learning architecture and the adopted blockchain. Other key evaluation criteria include whether a smart contract has been developed, whether it has been implemented in an IoT ecosystem, whether it performs an intrusion detection function, and finally whether CTI has been considered. Overall, while many papers followed a federated learning scenario, the majority have selected a standard (non-hierarchical) architecture, making their framework prone to its aforementioned limitations. Moreover, the global server functionality has been proposed to be hosted on a blockchain, which is a great initiative to attract the benefits of secure blockchain networking. However, most of the work has adopted a public blockchain network such as Ethereum, which leads to a breach of data privacy, as the updated model weights are shared publicly and can often be reversed to raw data using advanced adversarial attack techniques. A very small number of papers have used smart contract technology to automate the role of a global server, which is necessary for automated validation and control of the process. Finally, only one paper has aimed to provide a solution to organisations that are seeking to collaborate in the sharing of CTI for ML-based IDSs.

\section{Hierarchical Blockchain-based Federated Learning}
\label{fw}

\begin{figure}[!h]
  \centering
  \includegraphics[width=7.5cm, height=6cm]{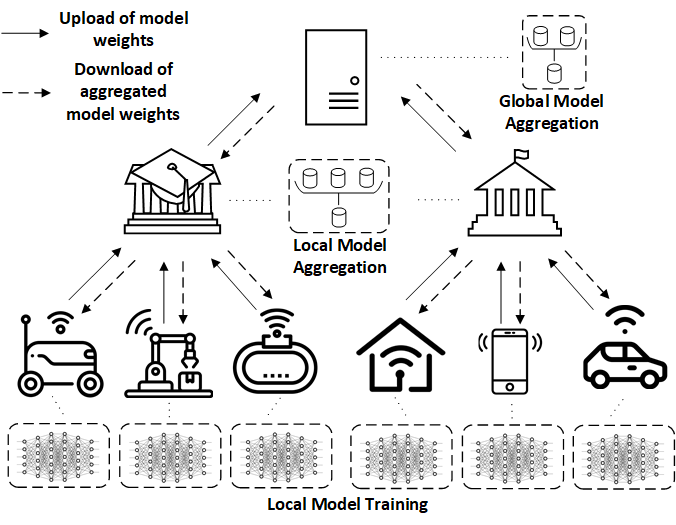}
  \caption{HBFL: architecture and data flow}
  \label{m}
\end{figure}

HBFL offers CTI to organisations in a secure, privacy-preserving, and automated manner. Participating organisations benefit from each other’s threat intelligence while ensuring the security and privacy of sensitive IoT endpoint data locally. Moreover, HBFL removes the trust requirement between organisations to mediate the orchestration process by the adoption of a smart contract running on a permissioned blockchain. The overall architecture used in the development of HBFL is presented in Figure \ref{m}. HBFL aims to design a universal ML-based IDS between organisations to improve detection performance. This is primarily due to the heterogeneity of the data generated and collected by each organisation's network and the vast differences in the operating environment incorporated by different IoT ecosystems. The learning model is exposed and trained on a multitude of heterogeneous data logs, each with a unique statistical distribution. The global IDS model correlates attack patterns over various organisational IoT networks. Therefore, the generalisability of ML-based IDSs is enhanced by an increase in the detection accuracy of locally unseen threats.

There are practical use cases for HBFL applications when an organisation $k$ aims to protect its IoT ecosystem $o$ using an ML-based IDS. Where $k$ collects a training data set $n$ locally from their network $o$; A) $k_1$ may not have collected/observed any malicious $m$ traffic for training in $o_1$.  For effective training of ML-based IDS, benign $b$ and malicious $m$ traffic are required as part of the training set $n$. Therefore, $k_1$ might depend on $m$ traffic generated in network $o_2$ available in another organisation $k_2$ for training. B) $k_1$ might have some $m$ traffic available for training, however, $n$ contains a limited number of $m$ that might not be sufficient in the design of an ML-based IDS to generalise across other $m$ scenarios. Which is a practical example where an organisation might not have experienced a wide range of attack traffic. Therefore, $k_2$ contribution is required to enrich $n$ with further $m$ traffic. In general, the collaboration between $k_1$ and $k_2$ to collect a common $n$ that contains a wider variety of $b$ and $m$ to improve the performance of ML-based IDS is a form of CTI.

\subsection{Privacy Preserving and Secure Learning}
A vertically integrated set of federated learning known as HFL has been chosen in the development of the HBFL framework. The key motivation for using an HFL architecture is the addition of servers in the middle layer between the top-level server and the IoT endpoints. In this framework, we refer to the top-level and middle-layer servers as reducers and combiners, respectively. The combiners are essentially the local orchestrating servers of an organisation. They reduce the centralisation of power in the reducer by providing further aggregation and control mechanisms locally to an individual organisation's IoT endpoints. This method attracts many benefits to the organisations participating in the HBFL operation. For instance, it overcomes the single-point-of-failure limitation in case of a reducer outage, as the learning process is maintained in each unaffected organisation separately because the clients can use the locally aggregated model by their respective combiners until it converges at the top-level's reducer. 

Moreover, HBFL supports large-scale deployments where a single reducer server is insufficient to communicate, obtain, and aggregate a large number of different model updates from the IoT endpoints, whether due to communication links or computation resource limitations. More fundamentally, for CTI sharing purposes, the model updates received directly from the IoT endpoints are preserved within the organisational perimeter (combiner) and never shared with any external entity; this acts as a defence mechanism against advanced adversarial attacks, such as reconstructing model updates to the training data used. It also avoids the necessity of differential privacy techniques which might degrade the detection accuracy of the IDS model \cite{bagdasaryan2019differential}. Furthermore, HBFL enables additional control points across the hierarchy which are used for defence and verification functions. For instance, each set of organisational weights is validated against adversary attempts at the combiner level, which consequently contains the effect of model and data poisoning techniques internally without reaching other organisational networks.

The aggregation in HBFL occurs at two different levels and different interval rounds; 1. Local aggregation - this is conducted by the combiner over a set of IoT endpoints as illustrated in Equation \ref{2}. 2. Global aggregation - this is conducted by the reducer over a set of participating organisations as illustrated in Equation \ref{3}. A key advantage of having a number of local weight aggregation rounds before a global aggregation is to obtain a locally converged and efficient model before the global aggregation with other organisational weights. This structure has the great performance benefit of decreasing the number of global aggregation rounds required to achieve global maximum performance. 
The local aggregated parameter $l$ from the combiner $k^{th}$ is given by
\begin{equation}
l_{c+1}^{k} = \sum\limits_{e_k=1}^{E_k} \frac{y_{e_k}}{y_k} w_{c+1}^{e_k} \hspace{1cm} for\ \ k=\{1,2,...,K\}
\label{2}
\end{equation}

where $c$ represents the local federated learning round, $k=\{1,2,...,K\}$ indicates the combiner, $y_{e_k}$ is the number of endpoints $e$ of the combiner $k$, $w_{c+1}^{e_k}$ are the model parameters trained from the endpoint $e$ of the combiner $k$, and $y_k$ indicates the total number of endpoints connected to the combiner $k$.
The aggregated global parameter $g$ from the reducer can be stated as
\begin{equation}
g_{t+1} = \sum\limits_{k=1}^{K} \frac{m_{k}}{m} l_{t+1}^{k}
\label{3}
\end{equation}

in which $t$ represents the number of global learning round, $m_k$ indicates the number of combiner, and $m$ is the total number of combiners.

%


The HBFL framework adopts a cloud-fog-edge topology where IoT endpoints and combiners are hosted in the edge and fog zones, respectively. The reducer functionality runs on a blockchain-based cloud network.  IoT endpoints are generally located on the edge of the organisational network ecosystem to assist in their functionality. The fog zone is a computing infrastructure to compute and store data between the edge and the cloud \cite{rahman2018semantic}. Cloud computing is the delivery of services and applications on flexible resources over the internet. The adopted topology reduces the costly communication with the cloud, with efficient local edge-fog updates. Therefore, the costs of securing, exchanging, and transmitting the communication between organisations and the cloud are reduced. The benefits of this topology are further clarified in Section \ref{dg} according to each component.


\subsection{Blockchain-based Orchestration and Conformance}
The adoption of a permissioned blockchain network in HBFL adds another layer of a  privacy-preserving mechanism. Only the reducer functionality will be deployed on the blockchain. The entire HBFL process between organisations will be orchestrated by a network of nodes on a blockchain. Therefore, we omit the trust of a single entity performing the global server roles, as well as the single-point-of-failure limitation. The identity management of each participating organisation is verified using a digital certificate managed and enforced by a smart contract upon registration and agreement. Through RBAC, each organisation will obtain limited read/write access to upload the locally aggregated parameters and collect the updated globally aggregated parameters for its use. 

The reducer in the HBFL architecture aggregates processed weights rather than individual IoT endpoint updates, thus, the necessity of secure aggregation and encryption is reduced at this level. Furthermore, the permissioned blockchain uses asymmetric cryptography to preserve communication between organisations. Therefore, the authenticity of the participant is verified through the use of private and public keys. The updated model weights received from each organisation are stored inside the blocks in the form of transactions. Since the transactions on blockchains are immutable and contain the source organisation of each record, adversarial attempts by an organisation to degrade the detection performance of the global HBFL model will be identified and accounted for.

Smart contracts are developed on the permissioned blockchain to automate the reducer role by monitoring, logging, and executing HBFL tasks. The smart contract contains the planned interactions and will act as the trusted shared agreement between the organisations. This will be a key motivation for participation due to the omission of the need to trust the involved parties, which is often a roadblock in collaborative processes. The smart contract will maintain a closely monitored HBFL environment. The smart contract relies on blockchains to automatically record the transactional information of the data or message flows that occurred while performing a process. Smart contracts not only check if interactions correspond to the agreed process, but also act as a validator on the communication details between the participant organisations and the reducer. The task is run through a preset structure of user-defined code based on the agreements and conditions set while designing the smart contract. 

By applying smart contracts to monitor the HBFL process, we ensure that only conforming interactions advance the state of the operation, and a tamper-proof and time-stamped audit log is kept. Initially, the smart contract will enrol an organisation by providing a digital certificate to be used in uploading the locally aggregated parameters. Parameters will be scanned for anomalies and adversary attempts prior to global aggregation. Once the required number of healthy parameters is received from the participants, global aggregation will occur. Finally, the smart contract will send the globally federated model parameters back to each organisation and terminate the HBFL process. Using smart contracts to facilitate collaboration between each participating organisation will enhance the HBFL process with two main functionalities;

\begin{itemize}
        \item \textbf{Active Mediation} - coordinates the collaborative HBFL process by controlling the start and end of each set of tasks and defined in the smart contract.
        
        \item \textbf{Monitoring} - records the execution status of the HBFL process and ensures that the interactions and message flows are conducted according to the agreed process conditions.

\end{itemize}

\subsection{Implementation}
\label{dg}
In this section, we demonstrate how the three main entity groups' functionalities are implemented, i.e., reducer, combiner, and endpoint. Reducer roles are performed by the smart contract on the permissioned blockchain. However, the remaining operational roles completed by the combiners and endpoints will be triggered by the smart contract and executed off-chain. The exit status of the off-chain processes will be sent back to the blockchain, which will be used for accounting, monitoring, and initiating the following process defined in the smart contract. There are four basic operating types of the following forms defined in the HBFL process;

\begin{itemize}
    \item \textbf{Start($Task_i$)} - The start status of task $Task_i$ is indicated by this message. If a task is meant to be performed outside the blockchain system, then the message shall be sent to the respective participants.
    
    \item \textbf{Hold}([$Task_i$.completed \dots $Task_n$.completed]) - If a task $Task_i$, and $Task_n$ are meant to be completed together to progress the workflow, then the obligation to perform these alternative tasks is indicated by this message. This is used before the aggregation process, where a number of updated parameters must be received.
    
    \item \textbf{End($Task_i$)} - The end status of task $Task_i$ is indicated by this message. If a task was performed outside the blockchain system, then the message shall be returned by the respective participants.
    
    \item \textbf{Monitor($Task_i$)} - If task $Task_i$ is complete, then the monitoring status of this task is indicated by this message. The \textit{Monitor} function ensures that the HBFL processes are executed according to the rules and conditions set within the smart contract and agreed upon by the participating organisations. The \textit{Monitor} records violations made while executing a process and terminates the HBFL process.
\end{itemize}

\subsubsection{Reducer}
The reducer is the single fundamental entity in the HBFL operation. It is hosted on a cloud-decentralised blockchain; therefore, each of the following tasks is automated and runs periodically directly using the smart contract. The cloud server can access a larger amount of data with excessive communication overhead and long latency, which will handle a large number of participating organisations (combiners). We assume a dummy initialising the reducer to orchestrate the HBFL until completion. There are two tasks required to be performed by the reducer; 1) \textbf{$Task_a$} - select a set of active organisations to participate, 2) \textbf{$Task_b$} - aggregate the received local model parameters $l$ from the combiners, and send the updated global parameters $g$ back to each organisation. In Algorithm \ref{alg2}, the reducer roles are defined. The reducer aggregates the locally aggregated parameters $l$ received from combiners $K$ indexed by $k$. The FedAVG aggregation technique is applied, where $t$ is the global federated learning round, and $m$ is the number of active combiners. In addition, the smart contract monitors the tasks executed on-chain by the reducer and off-chain by the combiner to ensure their conformance according to the agreed criteria. The final aggregated weights reflect the global CTI, which will be used in ML-based IDS deployment before the HBFL process is complete. 

\begin{algorithm}[!h]\small
  \SetKwData{Left}{left}
  \SetKwData{This}{this}
  \SetKwData{Up}{up}
  \SetKwFunction{Union}{Union}
  \SetKwFunction{FindCompress}{FindCompress}
 \textbf{Reducer} process initiated\\
$\textit{Start}$($Task_a$)\\
\Indp Select active and participating organisations\\
Initiate \textbf{Combiner} process \\
\Indm \textit{End}($Task_a$)\\
$\textit{Monitor}$($Task_a$) $\rightarrow$ Blockchain\\
$\textit{Monitor}$($Task_c$) $\rightarrow$ Blockchain\\  
 
$\textit{Start}$($Task_b$)\\
\Indp \textit{Hold}([$Task_{d_i}$.completed ... $Task_{d_n}$.completed])\\
$\textit{Monitor}$($Task_d$)\\
$S_{t} \leftarrow(\text {Set of } K \text {combiners) } $
     \For {each combiner $k \in S_{t}$ :}{
$g_{t+1}^{k} \leftarrow \text { Local update } (k, l_{t}) $

 $g_{t+1} \leftarrow \sum\limits_{k=1}^{K} \frac{m_{k}}{m} l_{t+1}^{k}$}
Send global model parameters $g$ to combiners\\
\Indm \textit{End}($Task_b$)\\
$\textit{Monitor}$($Task_b$)  $\rightarrow$ Blockchain\\
$\textit{Monitor}$($Task_e$) $\rightarrow$ Blockchain\\ 
  End \textbf{Reducer} process
  
\caption{Reducer (Smart Contract)}
\label{alg2}
\Indp\Indpp
\end{algorithm}\DecMargin{1em} 

\subsubsection{Combiner}
The combiners are hosted off-chain in the fog (perimeter) zone of the organisation. The fog zone enjoys more efficient communication with clients, allowing for faster model updates and efficient communication links compared to cloud servers. The results of each task conducted on the combiner will be sent to the global server to be recorded and monitored on-chain and progress the workflow of the smart contract. Each combiner is connected to a group of IoT endpoints located within the organisation's ecosystem. For each of the active combiners, there are three tasks performed; 1) \textbf{$Task_c$} - select a set of active IoT endpoints to participate, 2) \textbf{$Task_d$} - aggregate the received model parameters $w$ from the endpoints, and send the aggregated local parameters $l$ to the global server 3) \textbf{$Task_e$} - send the updated global parameters $g$ back to the endpoints. In Algorithm \ref{alg3}, the combiner role is triggered by the smart contract and is defined to be run off-chain in each organisation's network. The combiner selects a set of connected IoT endpoints $E$ indexed by $e$ to participate in the federated learning process. $E$ can be dynamically adjusted to the number of endpoints online with reliable communication conditions in a certain period. Each endpoint sends the trained model parameters $w$ to the combiner for aggregation. The FedAVG aggregation technique is applied, where $c$ is the local federated learning round, and $y$ is the number of active IoT endpoints. Local aggregated weights $l$ reflect an organisational intelligence and are sent to the reducer.

\begin{algorithm}[!h]\small
  \SetKwData{Left}{left}
  \SetKwData{This}{this}
  \SetKwData{Up}{up}
  \SetKwFunction{Union}{Union}
  \SetKwFunction{FindCompress}{FindCompress}
$\textit{Start}$($Task_c$)\\
\Indp Select active and participating IoT endpoints\\
Initiate \textbf{Endpoint} process \\

\Indm \textit{End}($Task_c$)\\

$\textit{Start}$($Task_d$)\\
\Indp \textit{Hold}([$Task_{f_i}$.completed ... $Task_{f_n}$.completed])\\

$S_{c} \leftarrow(\text {Set of } E \text { endpoints) } $
    
     \For {each endpoint $e \in S_{c}$ :}{
$l_{c+1}^{e} \leftarrow \text { Endpoint update } (e, w_{c}) $

 $l_{c+1} \leftarrow \sum\limits_{e=1}^{E} \frac{y_{e}}{y} w_{c+1}^{e}$}
Send aggregated local model parameters $l$ to reducer\\

\Indm \textit{End}($Task_d$)\\

$\textit{Start}$($Task_e$)\\
\Indp \textit{Hold}($Task_b$.completed)\\
Send aggregated global model parameters $g$ to endpoints\\
  \Indm \textit{End}($Task_e$)\\
  End \textbf{Combiner} process
  
\caption{Combiners}
\label{alg3}
\Indp\Indpp
\end{algorithm}\DecMargin{1em}

\subsubsection{Endpoint}
The IoT devices in this framework are defined as endpoints running in the edge zone due to their effectiveness when placed on the perimeter of IoT ecosystems. IoT devices collect the surrounding environmental data as a federated learning training data set. The functionality and integrity of the endpoints' operation are the responsibility of the parent organisation and are not monitored by the smart contract. This is a fair assumption, as organisations often have internal monitoring systems to detect faults and anomalies in IoT devices \cite{aboubakar2021review}. Additionally, an organisation can validate endpoint health by performing checks on its associated combiners; however, this is part of future work. In Algorithm \ref{alg4}, the roles of the endpoints are defined. There is one task performed by each of the participating endpoints; \textbf{$Task_f$} - train the ML model using local data samples, and send the updated parameters $w$ to the combiners. Once the participant is selected by the combiner, the ML model is trained using local data samples where $B$ is the local training batch size, $E$ is the number of local epochs, $\mathcal{P}$ is the local training set, $o$ is the loss of classification and $n$ is the local learning rate. The endpoints would preserve the privacy and security of their data logs by training the ML model locally and sending the updated parameters $w$ to the corresponding combiner, which reflects individual intelligence. Once the HBFL process is complete, the endpoints will receive the enhanced globally aggregated parameters $g$ from the combiner, which will be used for any practical intrusion detection. 

\begin{algorithm}[!h]\small
  \SetKwData{Left}{left}
  \SetKwData{This}{this}
  \SetKwData{Up}{up}
  \SetKwFunction{Union}{Union}
  \SetKwFunction{FindCompress}{FindCompress}
\Indp $\textit{Start}$($Task_f$)\\
\Indp $\mathcal{B} \leftarrow\left(\right.$ split $\mathcal{P}_{k}$ into batches of size $B$ ) \\
\For{each local epoch $i$ from 1 to $E$:}{
\For{batch $b \in \mathcal{B}$:}{
$w \leftarrow w- $n$ \nabla $o$ (w ; b)$ }}
Send updated model parameters $w$ to combiner\\
\Indm \textit{End}($Task_f$)\\
  End \textbf{Endpoint} process
\caption{Endpoints}
\label{alg4}
\end{algorithm}\DecMargin{1em}

\subsection{Security Analysis}
The HBFL architecture addresses several adversarial attacks associated with traditional federated learning techniques. HBFL provides a defence mechanism against free-riding and data/model poisoning attacks using the smart contract's \textit{monitor} component. Free riding attacks are where an intruder benefits from the global model parameters without contributing with legitimate data samples. Data or model poisoning adversaries \cite{liu2021privacy} aim to affect the global model to degrade its performance or make it perform in a certain way. In terms of IoT-IDS, the attacker can manipulate the global model to fail to detect certain attack techniques or tools which can be exploited in the future. The \textit{monitor} function of HBFL scans and verifies the parameters received from each organisation for any anomalies and discards them in the aggregation process in the case of an adversary. The exact technique of model weight anomaly detection such as k-means clustering will be covered in future work. In the case of a detected adversary, the smart contract will terminate the overall process by not sending the start command of the following task. Moreover, the HFL architecture adopted in HBFL naturally provides a segmentation feature between the organisations' networks, where it limits and contains the effect of an adversarial activity from spreading across all participant organisations, as it will only be contained within the adversary network.

The model weights received directly from the IoT endpoints are preserved within the organisational perimeter and never shared with any external entity; this acts as a defence mechanism against advanced adversarial attacks, reconstructing model updates to the training data used \cite{geiping2020inverting}. Furthermore, hosting the reducer on a permissioned blockchain reduces the possibility of attacking a central server in an untrusted environment. As only registered organisations with a unique ID can upload and verify the global model parameters, the operation is transparent and traceable if a participant uploads a forged parameter to the ledger. Therefore, any adversarial attempts by any organisation to degrade the detection performance of the global ML-based IDS model will be identified and accounted for. Furthermore, permissioned blockchains use a series of cryptographic algorithms, such as elliptic digital signatures and one-way cryptographic hash functions, to guarantee the authenticity and integrity of the data. Therefore, recorded transactions and model weights cannot be manipulated by an adversary. Overall, the proposed HBFL architecture provides a trusted source of data; monitored and controlled by a smart contract, providing secure and privacy-preserved collaboration between organisations.

\section{Experimental Evaluation}
\label{em}
An experimental use case has been implemented to evaluate the performance of the proposed HBFL framework for the detection of IoT intrusions. The feasibility and capabilities of the framework in the detection of several IoT attack scenarios are demonstrated. 

\subsection{Setup}

The ML model architecture used in the training and testing phases is a Deep Feed Forward (DFF). The architecture is chosen due to the two-dimensional tabular structure of the datasets. The data samples are fed forward through the model's various layers starting via an input layer that consists of a number of neurons equal to the same number of data features. There are four middle layers, each performing the Relu activation function. The final predictions are calculated in the output layer which consists of a single sigmoidal neuron. The neural units are initialised with randomly weighted connections which are optimised using the Adam algorithm during the training phase by mapping the input features to the desired output. The full list of parameters used in the design of HBFL is presented in Table \ref{par}. Although several parameter combinations were experimented with to achieve a reliable detection performance, the full exploration of hyper-parameter tuning is out of the scope of this paper. Therefore, it is highly likely to obtain better detection performance with an optimised set of parameters.

\begin{table}[h]\footnotesize
\centering
\caption{Federated training parameters}
\begin{tabular}{|l|l|}
\hline
\multicolumn{1}{|c|}{\textbf{Parameter}} & \multicolumn{1}{c|}{\textbf{Value}} \\ \hline
Hidden Layers                           & 4                                     \\ \hline
Neurons & 32, 16, 8, 4  \\ \hline
Activation Function                     & Relu                                  \\ \hline
Epochs                             & 10                                   \\ \hline
Batch Size                               & 128                                \\ \hline
Optimiser                          & Adam                                \\ \hline
Learning Rate                      & 0.001                               \\ \hline
Loss Function                            & Binary Crossentropy                 \\ \hline
Learning Rounds                & 10                                 \\ \hline

\end{tabular}
\label{par}
\end{table}

The performance of the ML model is evaluated using a set of standard evaluation metrics.The compound metrics are calculated based on the True Positive (TP), True Negative (TN), False Positive (FP) and False Negative (FN) numbers. The metrics used in this paper are;
\begin{itemize}
    \item \textbf{Accuracy} - the percentage of correctly classified data samples = $\frac{TP+TN}{TP+FP+TN+FN} \times 100$
    \item \textbf{Detection Rate (DR) } - the percentage of correctly classified attack data samples = $\frac{TP}{TP+FN} \times 100 $
    \item \textbf{False Alarm Rate (FAR)} - the percentage of benign data samples incorrectly classified =  $\frac{FP}{FP+TN} \times 100$
    \item \textbf{F1 Score} - the harmonic mean of precision ($\frac{TP}{TP + FP}$) and DR = $2 \times \frac{DR\;\times \;Precision}{DR\; +\; Precision}$
\end{itemize}
\noindent Each data set has been divided into training and testing sets in a ratio of 70\% to 30\%, respectively. Evaluation metrics are collected over the test set multiple times and mean results are presented.

A recent and widely used IoT IDS data set is used to evaluate the designed model. The data set is called NF-BoT-IoT-v2 \cite{sarhan2021towards}. Unlike the popular NSL-KDD data set, the chosen data set has been recently released; therefore, it represents modern and relevant IoT intrusions. The data samples are presented using NetFlow v9 attributes. NetFlow is a de facto standard in network monitoring and analysis due to its practicality and widely scalable deployment. NF-BoT-IoT-v2 is a new data set released in 2021, which is a NetFlow representative of the BoT-IoT \cite{koroniotis2019towards} data set. The BoT-IoT data set was released in 2018 by the ACCS Cyber Range Lab.  IoT and non-IoT traffic were generated using the Node-red and Ostinato tools, respectively, and Tshark is used to capture network packets in their native PCAP format. The NF-BoT-IoT-v2 data set was generated by converting the PCAP files to NetFlow based features using the nprobe tool. It contains the same four attack groups as its parent data set, i.e., DDoS, DoS, Reconnaissance, and Theft.  The total number of data samples is 37,763,497, where the attack samples are 37,628,460 (99.64\%) and 135,037 (0.36\%) are benign samples.

Further reprocessing techniques have been applied for optimal ML operations. The flow identifier features, such as source/destination IP addresses and ports, have been removed to avoid learning bias towards the attacking and victim endpoints. Furthermore, the Min-Max scaler technique has been used to normalise all feature values between 0 and 1 by applying  \begin{equation}X_*=\frac{X-X_{min}}{X_{max}-X_{min}}\end{equation}
\noindent where $X\textsubscript{*}$ represents the final output value ranging from 0 to 1. $X$ is the original input value and $X\textsubscript{max}$ and $X\textsubscript{min}$ indicate the maximum and minimum values of each feature, respectively. The Min-Max scaler was applied to each endpoint separately.

\begin{table}[h]\footnotesize
\centering
\caption{Data Distribution}
\label{ac}
\begin{tabular}{|l|l|l|l|l|l|l|l|}
\hline
\textbf{Org.}                   & \textbf{Endpoint} & \textbf{DDoS}             & \textbf{Recon.}           & \textbf{DoS}              & \textbf{Theft}            & \textbf{Benign}           & \textbf{Set}           \\ \hline
\multirow{2}{*}{\textbf{$k_1$}} & C1                & \checkmark & \checkmark & X                         & X                         & \checkmark & \multirow{2}{*}{$n_1$} \\ \cline{2-7}
                                & C2                & \checkmark & \checkmark & X                         & X                         & \checkmark &                        \\ \hline
\multirow{2}{*}{\textbf{$k_2$}} & C3                & X                         & X                         & \checkmark & \checkmark & \checkmark & \multirow{2}{*}{$n_2$} \\ \cline{2-7}
                                & C4                & X                         & X                         & \checkmark & \checkmark & \checkmark &                        \\ \hline
\end{tabular}%
\end{table}

The experimental use case developed considers two organisations $k_1$ and $k_2$ with unique IoT environments. The data sets $n_1$ and $n_2$ collected from each organisation contain a different set of attack classes, $m1 \subset n_1$ and $m2 \subset n_2$, where $m1 \neq m2$. This presents a realistic situation where each organisation have encountered and collected unique attack scenarios for ML training. Two different scenarios are considered in this paper; 1) a non-collaborative experiment where the models are trained on $n_1$ and evaluated on $n_2$ and vice versa using a traditional federated learning technique, 2) $k_1$ and $k_2$ collaborate using HBFL, and the global IDS model is trained and evaluated on both $n_1$ and $n_2$. The data set is split according to attack classes as demonstrated in Table \ref{ac}.

\subsection{Results}
\begin{figure*}[h!]
\centering
\begin{subfigure}{.49\textwidth}
  \centering
  \includegraphics[width=7.5cm, height=5cm]{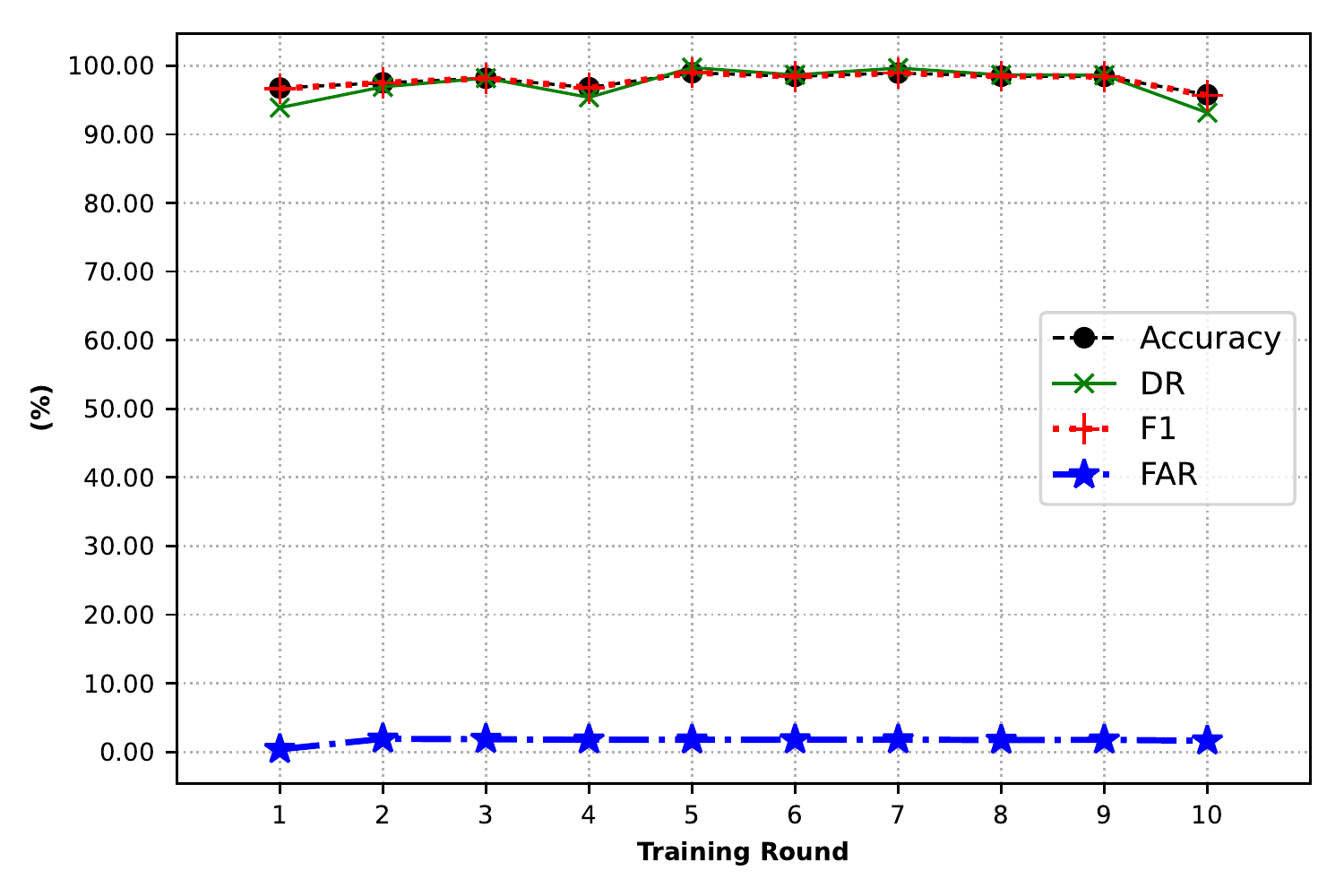}
  \caption{DoS}
\end{subfigure}
\hfill
\begin{subfigure}{.49\textwidth}
  \centering
  \includegraphics[width=7.5cm, height=5cm]{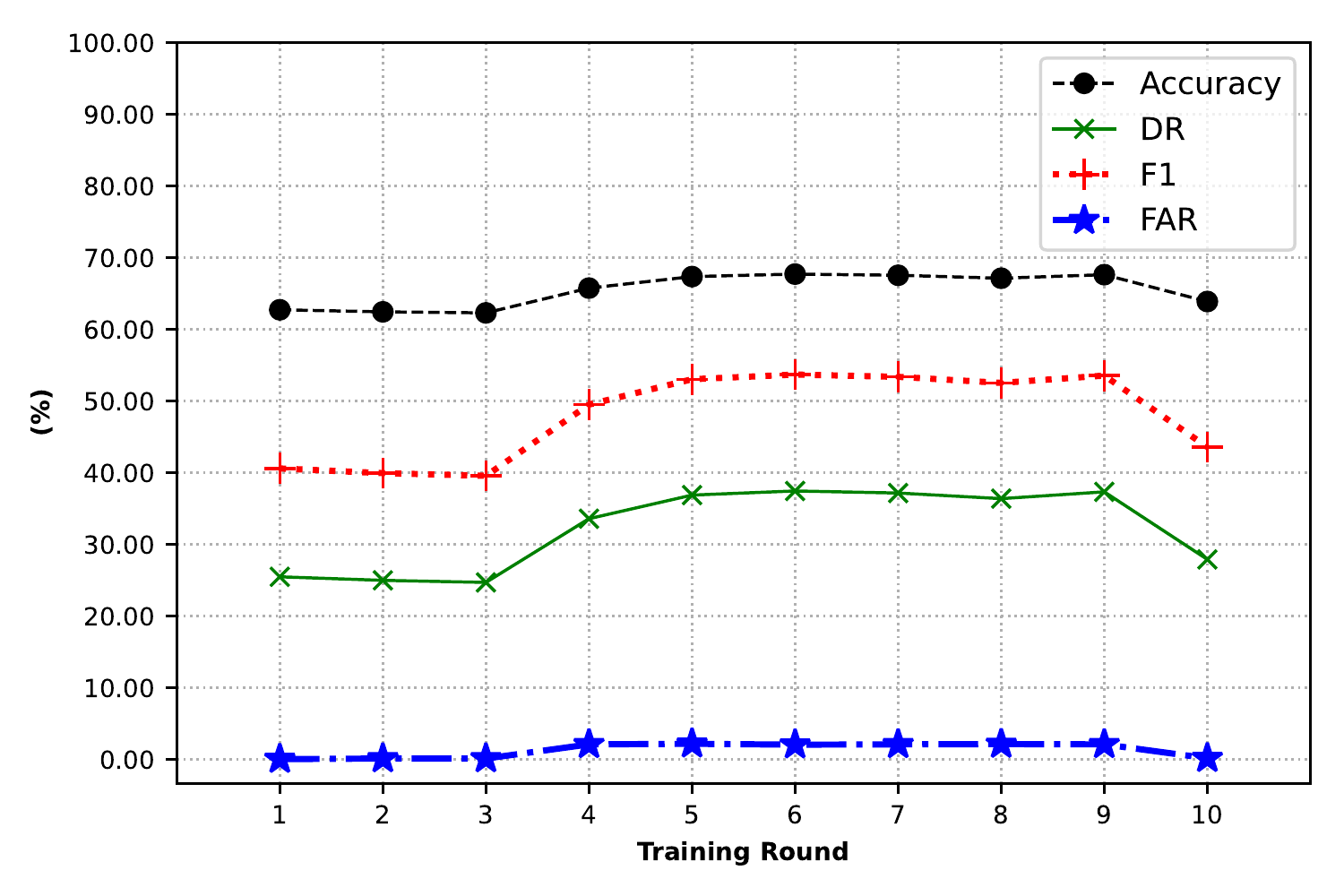} 
  \caption{Theft}
\end{subfigure}
\caption{Scenario 1 - Train on $n_1$, test on $n_2$}
\label{nc1}
\end{figure*}

\begin{figure*}[h!]
\centering
\begin{subfigure}{.49\textwidth}
  \centering
  \includegraphics[width=7.5cm, height=5cm]{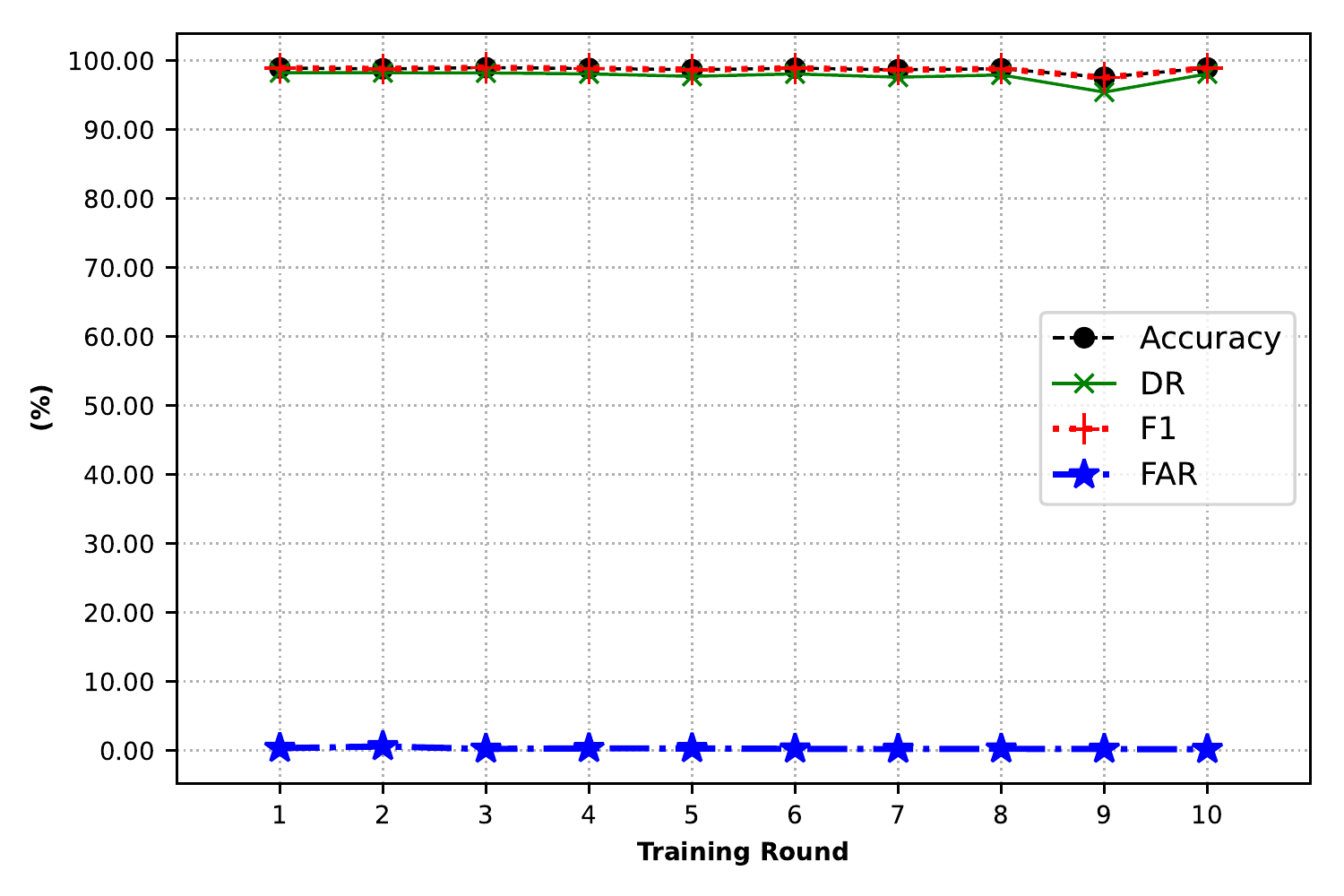}
  \caption{DDoS}
\end{subfigure}
\hfill
\begin{subfigure}{.49\textwidth}
  \centering
  \includegraphics[width=7.5cm, height=5cm]{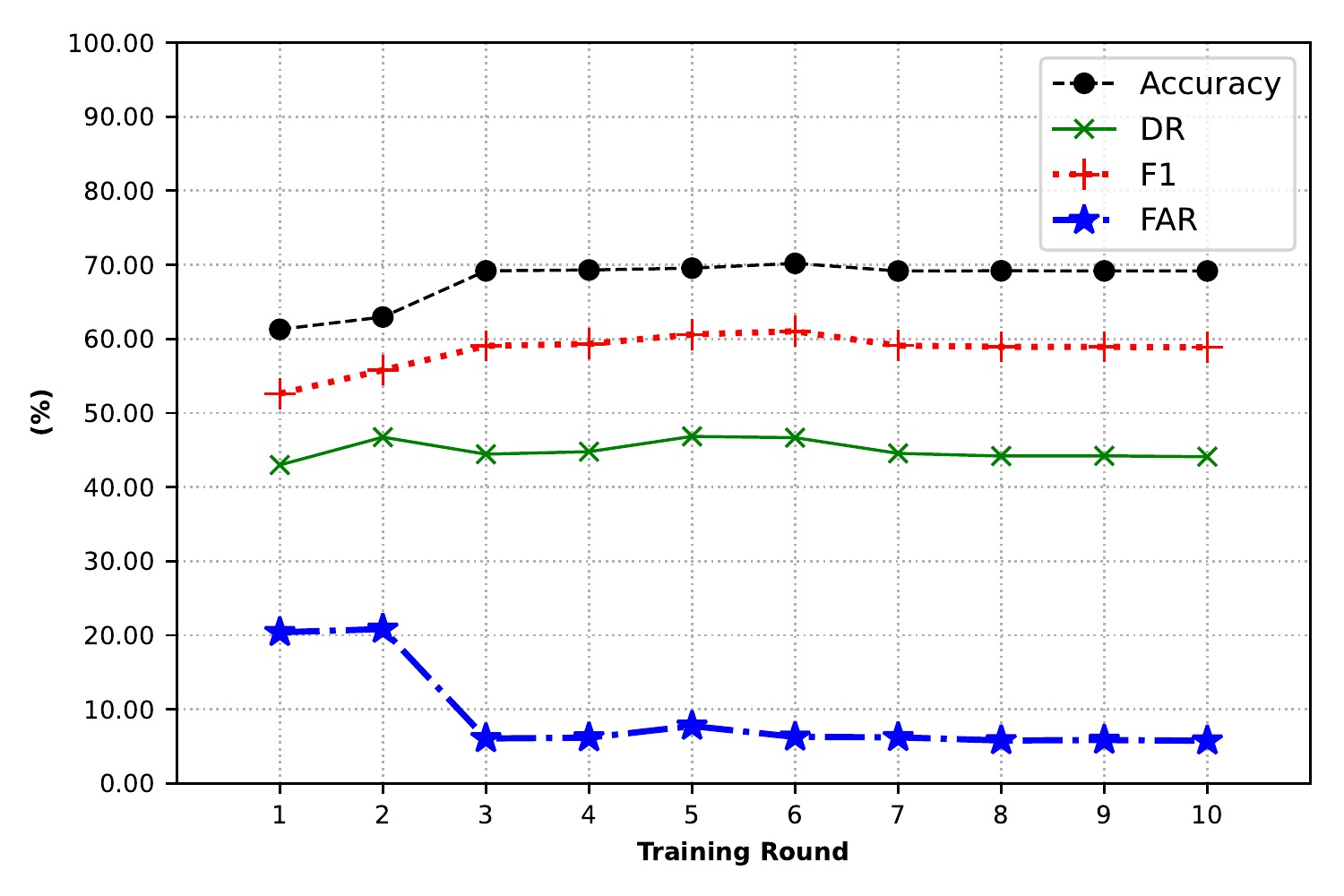} 
  \caption{Reconnaissance}
\end{subfigure}
\caption{Scenario 2 - Train on $n_2$, test on $n_1$}
\label{nc2}
\end{figure*}

\begin{figure*}[h!]
\centering
\begin{subfigure}{.49\textwidth}
  \centering
  \includegraphics[width=7.5cm, height=5cm]{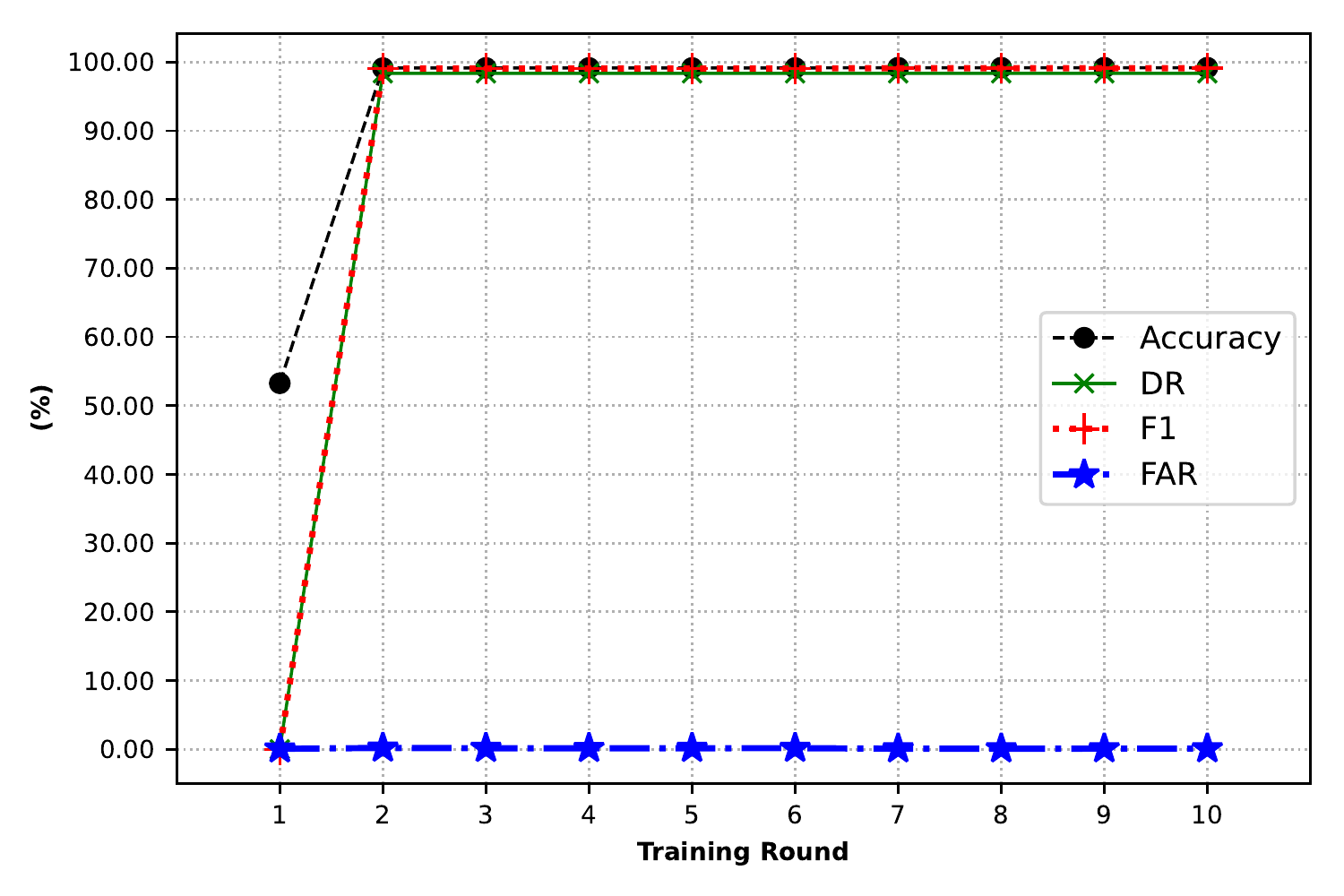}
  \caption{DDoS}
\end{subfigure}
\hfill
\begin{subfigure}{.49\textwidth}
  \centering
  \includegraphics[width=7.5cm, height=5cm]{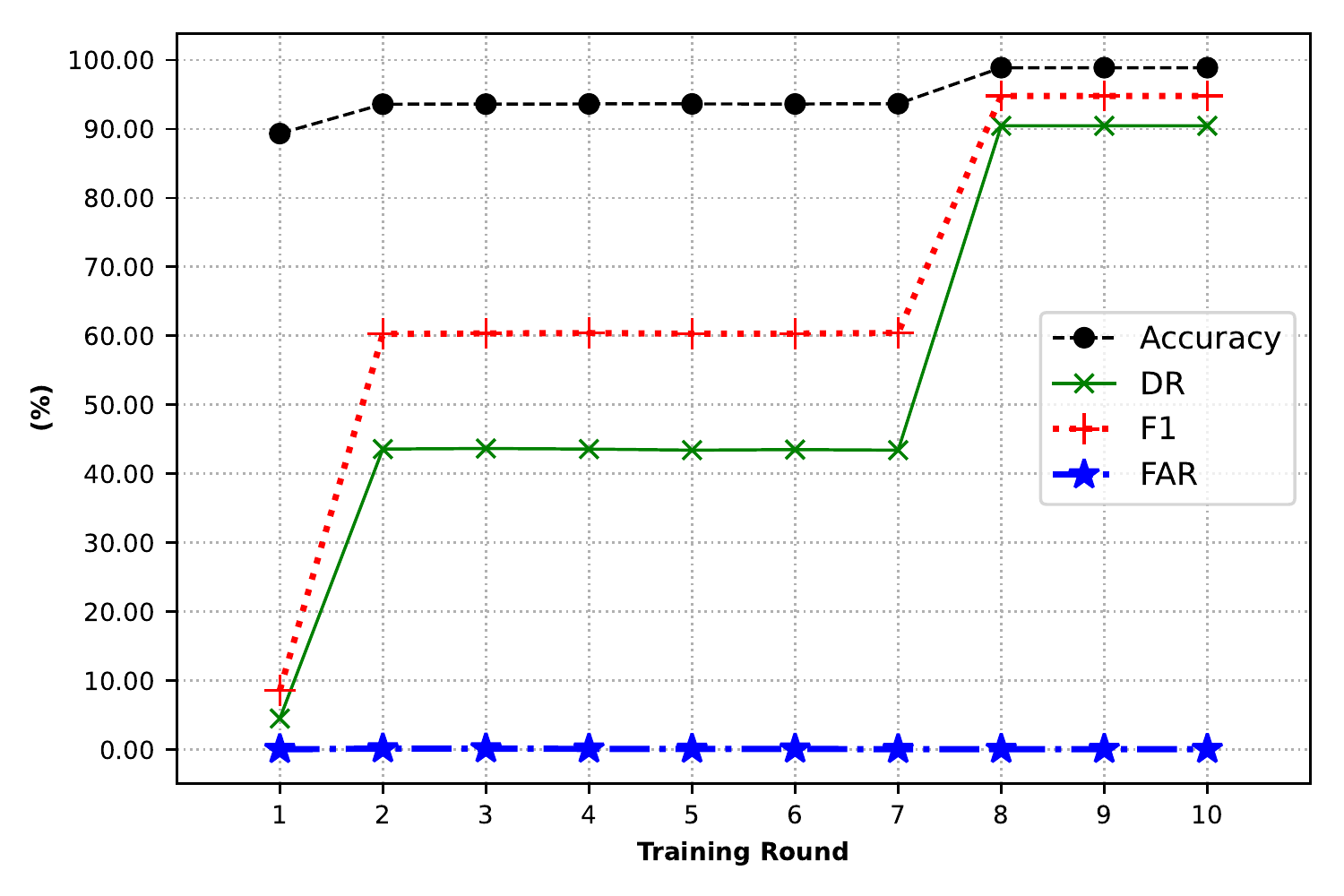} 
  \caption{Reconnaissance}
\end{subfigure}
\hfill
\begin{subfigure}{.49\textwidth}
  \centering
  \includegraphics[width=7.5cm, height=5cm]{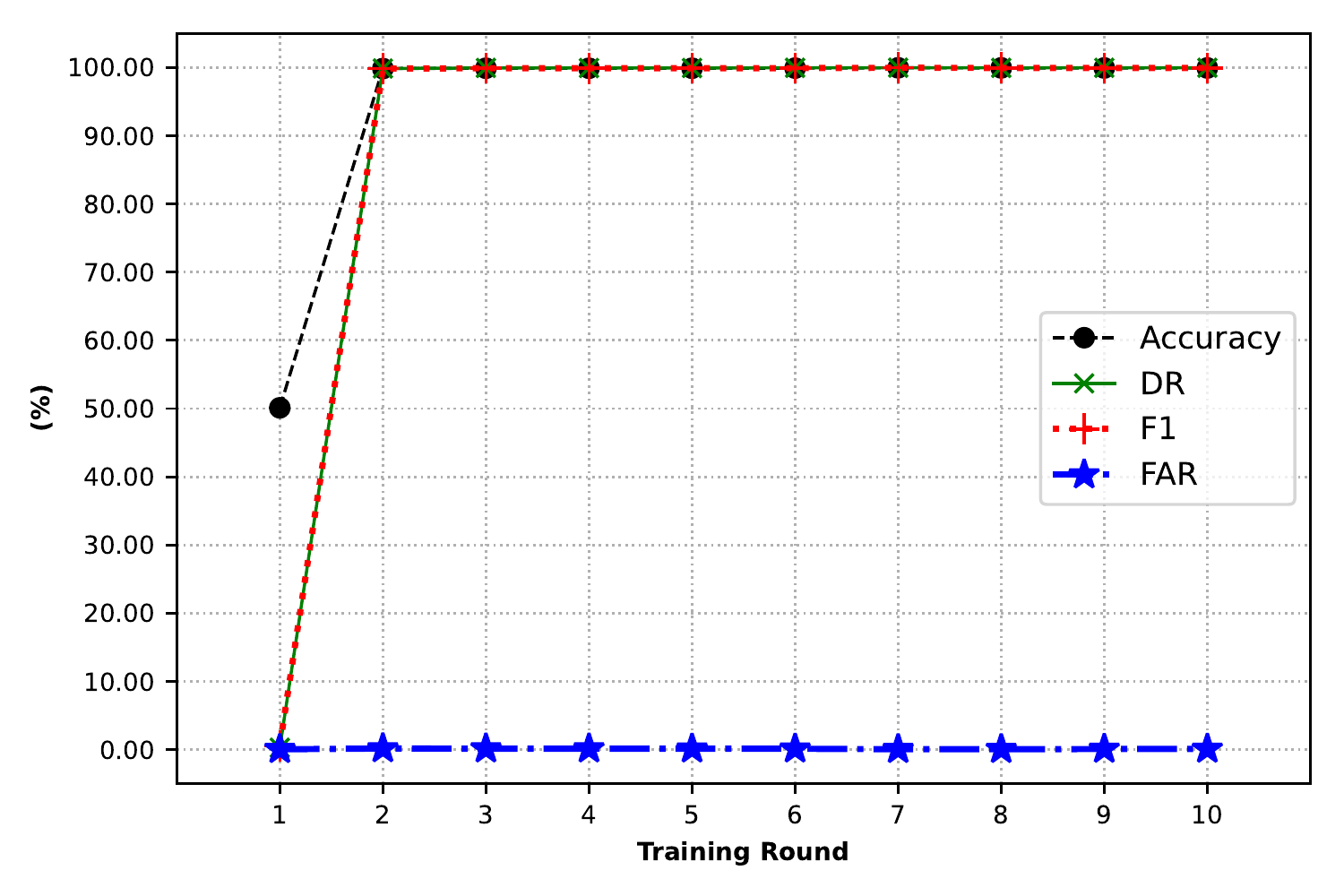}
  \caption{DoS}
\end{subfigure}
\hfill
\begin{subfigure}{.49\textwidth}
  \centering
  \includegraphics[width=7.5cm, height=5cm]{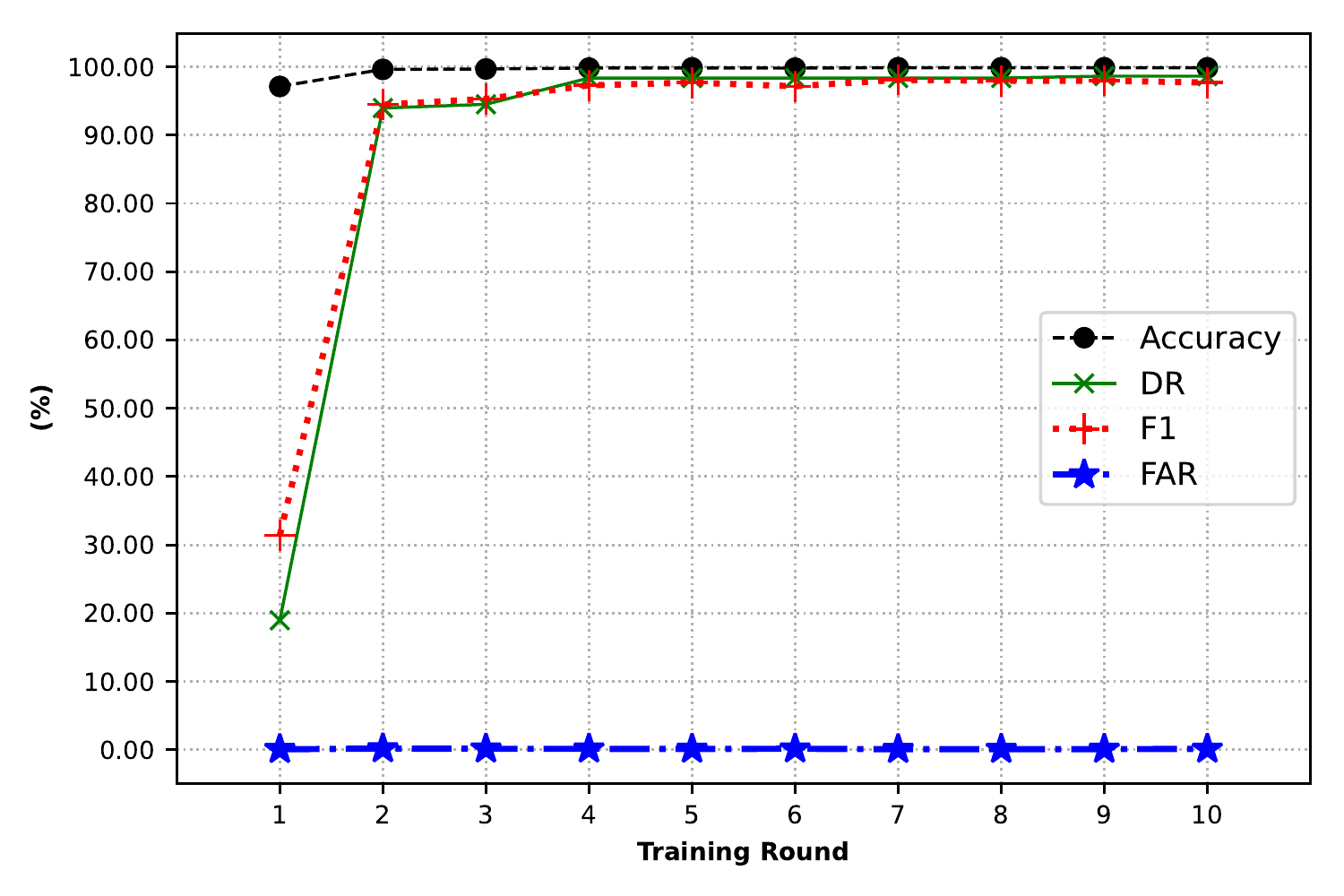} 
  \caption{Theft}
\end{subfigure}
\caption{Scenario 3 - HBFL}
\label{c}
\end{figure*}

The performance of HBFL is compared to the performance of a non-collaborative scenario. The results are demonstrated in Figures \ref{nc1}-\ref{c}, where each figure represents a different evaluation scenario. The sub-figures demonstrate the detection performance of each attack class used in the test set. The x-axis indicates the learning round, and the y-axis indicates the score of the evaluation metrics. Each line represents a different evaluation metric, where the results are calculated after each learning round to monitor the performance of the IDS model throughout the process. 

In Scenario 1, we assume a non-collaborative scenario, where organisation $k_1$ designs an IDS model on its local data set $n_1$. The trained model is evaluated on the data set $n_2$ of another organisation $k_2$ that contains different attack classes. The performance of the trained IDS model after each round is demonstrated in Figure \ref{nc1}. The final results indicate reliable DoS attack detection; however, poor theft attack detection with an accuracy of 95.77\% and 63.88\%, respectively. The further training rounds conducted did not have a significant impact as the detection performance stabilised after the 5$^{th}$ round.

In Scenario 2, we assume the opposite case of Scenario 1, where an IDS model is trained using the organisation $k_2$ data set $n_2$ and evaluated in the organisation $k_1$ data set $n_1$. The performance of the trained IDS model after each round is demonstrated in Figure \ref{nc2}. Again, the learning model cannot efficiently distinguish between the attack and benign network traffic in one of the attack classes. The final f1 scores of the DDoS and reconnaissance attack detections are 98.94\% and 58.85\%, respectively. The DDoS detection accuracy required one training round to reach the global maxima, where as three rounds were required for the reconnaissance attack.

In Scenario 3, we implement HBFL, where $k_1$ and $k_2$ collaborate in the design and training of a global IDS model by sharing local CTI. The model is trained and evaluated on $n_1$ and $n_2$ attack classes and the performances are demonstrated in Figure \ref{c}. As illustrated, all attack classes are nearly fully detected, and the FAR is minimal. Performance increases significantly after one round of training and stabilises fairly afterwards. However, efficient detection of the reconnaissance attack samples required eight training rounds to reach a maximum detection accuracy of 90.46\%. 

The numerical results of each scenario collected after the 10$^{th}$ federated learning rounds are presented in Table \ref{table}. Both Scenarios 1 and 2 achieve similar results where one of the unseen attack classes in the test data set is unreliably detected. The mean detection rate is 60.53\% and 71.1\% in Scenarios 1 and 2, respectively. However, in Scenario 3 (HBFL) the mean accuracy of each attack class is 99.71\%. A lower FAR is also observed in HBFL demonstrating the superiority of HBFL compared to other scenarios.

\begin{table}[h]\footnotesize
\centering
\caption{Model detection performance}
\label{table}
\begin{tabular}{|l|l|l|l|l|l|}
\hline
\textbf{Scenario}           & \textbf{Attack} & \textbf{Accuracy} & \textbf{DR} & \textbf{F1} & \textbf{FAR} \\ \hline
\multirow{2}{*}{\textbf{1}} & DoS             & 95.77             & 93.15       & 95.65       & 1.62         \\ \cline{2-6} 
                            & Theft           & 63.88             & 27.90       & 43.57       & 0.16         \\ \hline
\multirow{2}{*}{\textbf{2}} & DDoS            & 98.95             & 98.10       & 98.94       & 0.20         \\ \cline{2-6} 
                            & Recon.          & 69.17             & 44.09       & 58.85       & 5.75         \\ \hline
\multirow{4}{*}{\textbf{3 - HBFL}} & DoS             & 99.93             & 99.96       & 99.93       & 0.11         \\ \cline{2-6} 
                            & Theft           & 99.84             & 98.63       & 97.69       & 0.12         \\ \cline{2-6} 
                            & DDoS            & 99.18             & 98.37       & 99.12       & 0.10         \\ \cline{2-6} 
                            & Recon.          & 98.89             & 90.46       & 94.80       & 0.05         \\ \hline
\end{tabular}%
\end{table}

\subsection{Discussion}

Three scenarios were implemented and evaluated to compare the performance of IDS models enabled with CTI and non-CTI. The non-CTI enabled models (Scenario 1 and 2) are implemented using traditional federated learning techniques used widely in the literature. The test sets for Scenarios 1 and 2 are extracted from the unseen attack classes demonstrating the non-shared intelligence from other organisations. The CTI enabled model (Scenario 3) is implemented using the proposed HBFL architecture to enforce the additional privacy and security control points. The experimental results demonstrate the extreme superiority of HBFL and the importance of CTI compared to other scenarios. In Scenarios 1 and 2, one of the unseen attack classes, theft and reconnaissance, respectively are unreliably detected through out the experiment. 

DDoS and DoS attacks were fully detected as unseen attacks. This can be explained by the similar nature of the attack classes and the techniques used to carry out the activities. Therefore, the model is able to successfully transfer the knowledge learnt from one attack class to the other. In general, the lack of CTI between organisations presents a serious cyber risk to ML-based protected networks in the detection of unseen attack classes. On the other hand, the HBFL global model is capable of efficiently detecting the full set of attack classes and threat intelligence shared by each participant organisation. This also confirms that the additional privacy and security mechanisms provided by HBFL do not have an impact on the accuracy of intrusion detection. 

\section{Conclusion}
In this paper, we propose a novel CTI enabling framework known as HBFL for ML-based IDSs. HBFL aims to strengthen the organisational IoT security posture by providing insights collected from different organisational environments. The framework adopts a cloud-fog-edge topology in which the IoT endpoints, combiners, and reducer are hosted in the edges, fog, and cloud, respectively. A secure architecture was designed by integrating HFL and permissioned blockchains. In addition, a smart contract has been developed to ensure the conformance of executed tasks according to pre-defined acceptance criteria. We have tested our solution and demonstrated its feasibility by implementing it and evaluating intrusion detection performance using a key data set. The outcome of the framework is a robust, high performance, and securely designed ML-based IDS to protect and preserve the integrity of IoT networks. Future work includes implementing an adversary detection model using clustering of k-means to implement in combiners.


\bibliography{main.bib}

\end{document}